\documentclass[aps,prd,twocolumn,showpacs]{revtex4}
\usepackage{graphicx}
\usepackage{latexsym}
\usepackage{amsmath}
\usepackage{amsfonts}
\usepackage{amssymb}
\usepackage{verbatim}

\newcommand{\be}{\begin{equation}}
\newcommand{\ee}{\end{equation}}
\newcommand{\bea}{\begin{eqnarray}}
\newcommand{\eea}{\end{eqnarray}}

\newcommand{\pa}{\partial}

\newcommand{\bb}{\bibitem}
\def\pls{\partial\!\!\!/}
\def\bb{\bibitem}
\def\as{A\!\!\!/}

\def\ps{p\!\!\!/}
\def\bs{b\!\!\!/}

\def\ds{\partial\!\!\!/}

\def\bb{\bibitem}
\newcommand{\ben}{\begin{eqnarray}}
\newcommand{\een}{\end{eqnarray}}

\begin{document}

\title{Induced Chern--Simons-like action in Lorentz-violating massless QED}
\author{$^{a}$F.A. Brito, $^{b}$L.S. Grigorio, $^{b}$M.S. Guimaraes, $^{a,c}$E. Passos,
$^{b}$C. Wotzasek} \affiliation{$^{a}$Departamento de F\'\i sica,
Universidade Federal de Campina Grande, Caixa Postal 10071,
58109-970  Campina Grande, Para\'\i ba,
Brazil\\
$^{b}$Instituto de F\'\i sica, Universidade Federal do Rio de Janeiro,
\\Caixa Postal 21945, Rio de Janeiro, Brazil
\\
$^{c}$Departamento de F\'\i sica, Universidade Federal da Para\'\i
ba, Caixa Postal 5008, 58051-970 Jo\~ao Pessoa, Para\'\i ba, Brazil}


\begin{abstract}In the present work, we study different aspects of
Lorentz and CPT symmetry violation in extended massless QED. By
following the observation that the 2+1 dimensional
Maxwell--Chern--Simons theory can be originated from the 3+1
dimensional Chern--Simons-like action, we also focus on the fermion
sector to relate the 3+1 dimensional extended massless QED to  2+1
dimensional massive QED. We take advantage of this to state that the
Chern--Simons-like action in extended massless QED can be induced
with its coefficient being well defined and finite just as its 2+1
counterpart. We make use of three different regularization schemes
by inducting the Chern--Simons-like in 3+1 dimensions to support the
conjecture.
\end{abstract}
\pacs{11.30.Er, 11.10.Wx, 12.20.Ds} \maketitle


\section{Introduction}
The main goal of this investigation is to revisit the question posed
by the ambiguity problem of the Chern--Simons-like action induced by
radiative corrections in the photon sector of the Lorentz-violating
extended Standard Model \cite{kostelecky}. Several theoretical
investigations have pointed out the possibility of the Lorentz and
CPT symmetries being violated in nature
\cite{kostelecky,0,1,2,4,5,6,7,biref,8,9,10,11,11.2,12,13,14}.  These
symmetries would then be approximately realized. A great body of
phenomenological results was obtained by realizing that this
violation can indeed be generated in the photon sector of QED by
adding the Chern--Simons-like action ${\cal
L}^{3+1}_{CS}=\frac{1}{2}k_\mu\epsilon^{\mu\alpha\beta\gamma}F_{\alpha\beta}A_\gamma$,
with $k_{\mu}$ being a constant quadrivector characterizing a
preferred direction of the space-time. As one knows, this extended
electromagnetism does not break the gauge symmetry of the action and
equations of motion but it does modify the dispersion relations for
different polarization of photons. The Chern--Simons-like action is
known to have some important implications, such as birefringence of
light in the vacuum \cite{biref}. Many interesting investigations in
the context of Lorentz-CPT violation have appeared recently in the
literature. For instance, several issues were addressed, such as
\v{C}erenkov-type mechanism called ``vacuum \v{C}erenkov radiation''
to test the Lorentz symmetry \cite{ptl}, changing of gravitational
redshifts for differently polarized Maxwell--Chern--Simons photons
\cite{kkl}, evidence for the Lorentz-CPT violation from the
measurement of CMB polarization \cite{bjx}, supersymmetric
extensions \cite{bch}, breaking of the Lorentz group down to the
little group associated with $k_\mu$ \cite{hle} and magnetic
monopoles inducing electric current \cite{brz}.

A possible modification in the fermionic sector is to consider the CPT-odd term
$\bar{\psi}{\bs}\gamma_5\psi$ with a constant quadrivector $b_\mu$
introducing CPT symmetry breaking.  This modification brings one of the most interesting and
controversial results regarding the dynamical origin of the parameter
$k_\mu$  present in the Lorentz and CPT symmetry breaking
obtained when we integrate over the fermion fields in the modified
Dirac action. The result is the induction of the Chern--Simons-like action
via radiative corrections which may lead to a relation between
$k_\mu$ and $b_\mu$, such that the last is given as an input in the theory.

The induction of the Chern--Simons-like action, ${\cal L}_{CS}$, is
an important result in the study of the Lorentz symmetry violation
\cite{2,kostelecky,7}. This term, naturally emerges as a perturbative
correction in the theory suggested in \cite{kostelecky} as a possible
extension of QED by an axial-vector term \ben\label{QEDext} {\cal
L}=\bar{\psi}( i \pls - M )\psi - \bar{\psi} \bs \gamma_{5}\psi -
e\bar{\psi} \as \psi. \een By carrying out the integration over
fermions it is possible to obtain a relation between the
coefficients $k_{\mu}$ and $b_{\mu}$ in terms of loop integrals as
$k_{\mu}=C b_{\mu}$ with some of them being possibly divergent.
Therefore since one has to implement a proper regularization to
calculate these integrals, the constant $C$ relating the
coefficients turns out to be dependent on the regularization scheme
used \cite{bpp}. The ambiguity of the results, manifested in the
dependence on the regularization scheme, has been intensively
discussed in the literature. Several studies have shown that $C$ can
be found to be finite but undetermined \cite{23, 24, 25, 26, 27}. {
Astrophysical observations impose very stringent experimental bounds
on $k_{\mu}$ suggesting that it should vanish. Since the coefficient
$k_{\mu}$ of the radiatively induced Chern--Simons-like action
depends on $b_\mu$ it is natural to expect that the constant $b_\mu$
can also suffer an experimental bound in this framework. However, if
ambiguities are present there is no way to know for sure to what
extent the experimental bounds would constrain the constant $b_\mu$,
simply because $C$ is undetermined. In other words, we cannot define
the fate of the constant that is responsible for the Lorentz and CPT
violation in the fermion sector by simply measuring $k_\mu$. In the
following we are going to reduce the well studied Lagrangian
(\ref{QEDext}) to the massless case in attempting to shed some light on the issue of
inducing the Chern--Simons-like action with no ambiguities.}

The paper is organized as follows. In Sec.~\ref{sec1}, we show how
to relate the 3+1 dimensional extended {\it massless} QED to 2+1
dimensional {\it massive} QED. We also comment on the relation
between the induced 3+1 dimensional Chern--Simons-like term and its
counterpart in 2+1 dimensions. In Sec.~\ref{sec2}, we show how to
induce the Chern--Simons-like term in an extended massless QED. In
Sec.~\ref{sec4}, we make use of the dimensional regularization and
the momentum cut-off regularization. In Sec.~\ref{temp}, we consider
the induction of the Chern--Simons-like term at finite temperature.
In this section, we use the limit of zero temperature as our third
regularization scheme. Finally, in Sec.~\ref{conc} we present our
ending comments.

\section{$3+1$ Lorentz violation as the origin of $2+1$ fermionic mass}
\label{sec1}

In this section we shall point an interesting motivation to study
the 3+1 dimensional massless QED. The approach we intend to take in
the following is based on the well known fact that in $2+1$ QED the
massive fermionic degrees of freedom gives rise to the CS term
without any ambiguity in its coefficient. It is useful to review
this argument here. The physical origin of this phenomenon is the
fact that the fermionic mass in $2+1$ dimensions is just a spin
density, that is, massive fermions have definite spin in this
dimensionality with a fixed orientation given by the sign of the
mass term. As result $P$ and $T$ symmetries are broken. So, for
instance, if we are interested in a scale well bellow that defined
by the fermionic mass, fermions will only contribute through their
quantum fluctuations disturbing the electromagnetic propagation. The
resulting effective theory must convey the information of the $P$
and $T$ discrete symmetry breaking and hence the induction of the CS
term. This term does not depend on the absolute value of the
fermionic mass, only on its sign, while all the other corrections
are attenuated by the mass which makes the resulting
Maxwell--Chern--Simons (MCS) a suitable effective field theory for
QED with very massive fermions.

What is important for the present work is to observe that the MCS
theory can be viewed as the dimensionally projected theory \cite{mm}
of the CPT-odd Carroll--Field--Jackiw model, which is just the
Maxwell theory augmented by the CPT-odd term discussed previously.
This simple observation prompt us to search for a system in $3+1$
dimensions having $2+1$ dimensional QED as its dimensional
reduction. An immediate point to call up to attention is that the
$P$ and $T$ symmetry breaking mass in $2+1$ dimensions bears no
relation to the $3+1$ dimensional fermionic mass. In fact it is
easily seen that the $2+1$ dimensional fermionic mass term is
related to the axial-vector term $\bar{\psi} \bs \gamma_{5}\psi$ in
$3+1$ dimensions. Choosing the $b_\mu$ as a space-like vector and in
the $3$-direction with $b_3=m$ and working in the chiral
representation for the gamma matrices
\begin{eqnarray}
\label{gamma}
&&\gamma^5 = \left( \begin{array}{cc}
-1 & 0 \\
0 & 1 \end{array} \right); \;\;\; \gamma^{\mu} = \left( \begin{array}{cc}
0 & \sigma^{\mu} \\
\bar{\sigma}^{\mu} & 0 \end{array} \right);\nonumber\\&& \sigma^{\mu} =(1, \overrightarrow{\sigma} );\;\;\; \bar{\sigma}^{\mu} =(1, -\overrightarrow{\sigma} )
\end{eqnarray}
we obtain
\begin{eqnarray}
\label{massproj}
b_{\mu}\bar{\psi} \gamma^{\mu}\gamma_{5}\psi \rightarrow m (\bar{\psi}_L \psi_L + \bar{\psi}_R \psi_R)
\end{eqnarray}
were $\psi_L$ and $\psi_R$ are the left and right two-component
spinor fields, respectively. This result is intuitively clear if we
realize that the $P$ and $T$ symmetry breaking generated by the
fermionic mass term in $2+1$ is just the dimensional reduction of
the Lorentz symmetry breaking in $3+1$ dimensions. The Lorentz
violating vector breaks the Lorentz symmetry in $3+1$ dimensions
because it defines a preferred direction and a orientation. Upon
fixing its direction and projecting the theory on the corresponding
orthogonal plane there remains the preferred orientation
encapsulated in the sign of $m$ in (\ref{massproj}) which is the
source of the $P$ and $T$ symmetry breaking in $2+1$ dimensions.

\begin{figure}[h]\centerline{\includegraphics[{angle=90,height=8.0cm,angle=270,width=8.0cm}]
{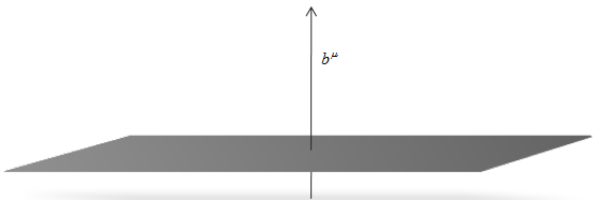}} \caption{The space like $b_\mu$-vector in the third
direction. The 2+1 dimensional action for the massive QED and
Chern--Simons term is on the plane.}\label{plane}
\end{figure}

The Dirac equation that follows from (\ref{QEDext}),
\begin{eqnarray}
\label{direqu}
( i \pls - M  - \bs \gamma_{5} - e \as) \psi = 0 \, ,
\end{eqnarray}
can be decomposed in its chiral components as
\begin{eqnarray}
\label{direquchi}
( i \sigma^{\mu} \partial_{\mu} - e \sigma^{\mu}A_{\mu} - \sigma^{\mu} b_{\mu}) \psi_{R} - M\psi_{L} = 0 \\
( i \bar{\sigma}^{\mu} \partial_{\mu} - e \bar{\sigma}^{\mu}A_{\mu} + \bar{\sigma}^{\mu} b_{\mu}) \psi_{L} - M\psi_{R} = 0.
\end{eqnarray}
For $M\rightarrow0$ these equations decouple as is well known since
chiral symmetry is restored. A `natural' setup considers a dynamical
dimensional reduction mechanism via {\it domain wall background}. In
this point of view $M\equiv m\mp m\tanh(\frac{x^3}{\Delta})\to0$
outside ($x^3\to\pm\infty$) the two-dimensional domain wall with
thickness $\Delta$, whereas $M\equiv m$ inside the domain wall at
$x^3\simeq0$. Thus, the 3+1 theory outside the domain wall has
chiral symmetry but breaks Lorentz and CPT symmetry, while in the
2+1 theory inside the domain wall the Lorentz and CPT symmetry is
restored but the chiral symmetry is broken. Similarly, in this
mechanism the induced 3+1 Chern--Simons-like term outside the domain
wall is expected to match the induced and determined Cherns--Simons
term in 2+1 dimensions inside the domain wall, since the latter can
be brought from the former by just decomposing the 3+1 dimensional
fields into 2+1 dimensional fields, i.e., $A^\mu\to(A^a,\varphi),\
k^\mu\to (k^a,-k_{3}),\ a=0,1,2$, such as ${\cal
L}_{CS}^{2+1}=\frac12 k_3\epsilon^{abc}F_{ab}A_c$, up to
scalar-vector field coupling. Thus, according to this match, one
expects that at least {\it one component} of the Chern--Simons-like
term in 3+1 dimensions should be finite and determined. What is
interesting though is that by further projecting into the plane
$x^{3} = 0$ and again working with a space like $b_\mu$-vector in
the $3$-direction with $b_3=m$
--- see Fig.~\ref{plane} --- the first equation in
(\ref{direquchi}), say, will read
\begin{eqnarray}
\label{direquchi3d} ( i \sigma^{a} \partial_{a} - e \sigma^{a}A_{a}
- e \sigma^{3}\phi - \sigma^{3} m) \psi_{R} = 0 .
\end{eqnarray}
where $\phi \equiv A_{3}$, $a=0,1,2$ and $\partial_3\psi=0$ (there
is no transversal momentum $p_3$). By choosing an appropriate
representation for the Dirac matrices in $2+1$ this equation may
be shown to follow from the $2+1$ action
\begin{eqnarray}
\label{qed3d} S=\int d^3 x \bar{\psi}_{R}( i \pls - e \as - e \phi -
m) \psi_{R}.
\end{eqnarray}
As already discussed, the action (\ref{qed3d}) is known to induce a
finite and well determined Chern--Simons action in $2+1$. Then, as
we anticipate by using a domain wall background point of view, it is
reasonable to expect that the proper fermionic action in $3+1$ that
will induce the CPT-odd term without ambiguity is given by
(\ref{QEDext}) with $M=0$, i.e., an extended massless QED. This is
precisely what happens, at least for the three different
regularization schemes used, as we shall show in the following
sections.

\section{The induced Chern--Simons action in an extended massless QED}
\label{sec2}

Let us introduce an extension of the usual Lagrangian density in the
absence of mass term \cite{8,12,Scarpelli:2008fw} as follows \ben {\cal
L}=\bar{\psi}( i \pls - \bs\gamma_{5} - e \as) \psi. \label{QEDQ}
\een The one loop effective action  $S_{eff}[b,A]$ of the gauge
field $A_{\mu}$ in this theory can be expressed in the form of the
following functional trace \be\label{Ef11} S_{eff}[b,A]=-i\,{\rm
Tr}\,\ln\big[\ps- \bs\gamma_{5} - e \as \big]. \ee Notice that this
point is delicate by the fact that the trace symbol--${\rm Tr}$
stands for the trace over Dirac matrices, trace over internal space,
as well as for the integration in momentum and coordinate spaces.
Hence, in this present case the Eq.(\ref{Ef11}) cannot be directly
calculated and written as $S_{eff}=\int d^{4}x {\cal L}$ since the
electromagnetic field-$A_{\mu}$ is coordinate dependent and,
therefore, does not commute with functions of momentum. In fact, it
is not an easy task to separate out the momentum and space dependent
quantities, and carry out the integrations in respective spaces. To
solve this problem, we will use the method of derivative expansion
\cite{de}(see also \cite{bpp}), and proceed as in the following.
This functional trace can be represented as \bea
S_{eff}[b,A]=S_{eff}[b]+S_{eff}^{\,\prime}[b,A], \eea where the
first term is $S_{eff}[b]=-i\,{\rm Tr}\ln\big[\ps - \bs\gamma_{5}]$,
which does not depend on the gauge field, and the only nontrivial
dynamics is governed by the second term $S_{eff}^{\,\prime}[b,A]$,
which is given by the following power series \be \label{ea}
S_{eff}^{\,\prime}[b,A]=i\,{\rm Tr} \sum_{n=1}^{\infty}\frac1n
\Biggl[\frac1{\ps - \bs\gamma_5}\,e \as\Biggr]^n. \ee To obtain the
Chern--Simons-like action we should expand this expression up to the
second order in the gauge field \be \label{Ef1}
S_{eff}^{\,\prime}[b,A]=S_{eff}^{(2)}[b,A]+\ldots \, . \ee The
ellipsis in (\ref{Ef1}) stands for the terms of higher orders in the
gauge field. Here \bea\label{efa2}
S_{eff}^{(2)}[b,A]=-\frac{ie^{2}}{2}{\rm
Tr}\bigl[S_b(p)\;\as\;S_b(p)\;\as\bigl], \eea where $S_b(p)$ is the
$b^{\mu}$-dependent propagator of the theory defined as
\bea\label{p1} S_b(p)=\frac{i}{ \ps-  \bs\gamma_5}. \eea Now, we can
use the identity \bea A_{\mu}(x)S_{b}(p)=S_{b}(p-i\pa)A_{\mu}(x),
\eea with its respective expansion of operators given in the form
\bea S_{b}(p-i\pa)=S_{b}(p)+ S_{b}(p)\,\ds\,S_{b}(p)+\cdot\cdot\cdot
\eea Substituting this back into the expression (\ref{efa2}), we
obtain \bea \label{efa2a} S_{eff}^{(2)}[b,A]=\int d^4x
\Pi^{\mu\lambda\nu}\pa_{\lambda}A_{\mu}A_{\nu}, \eea with the one
loop self-energy tensor being given by \bea \label{efa2aa}
\Pi^{\mu\lambda\nu}=-\frac{ie^2}{2}\int
\frac{d^4p}{(2\pi)^4}
{\rm
tr}\left[S_b(p)\gamma^{\mu}S_b(p)\gamma^{\lambda}S_b(p)\gamma^{\nu}
\right], \eea where {\rm tr}, means that the trace just acts over
the gamma matrices. After some algebraic manipulations, we can
rewrite the above propagator as \bea S_b(p)=W_{1}P_{L}+W_{2}P_{R},
\eea where \bea P_{R,L}=\frac{1\pm\gamma_{5}}{2}, \eea and \bea
W_{1}=\frac{i(p+b)^{2}(\ps-\bs)}{(p^{2}-b^{2})^{2}}\;\,\,,\,W_{2}=\frac{i(p-b)^{2}(\ps+\bs)}{(p^{2}-b^{2})^{2}}.
\eea Therefore, the expression to the self-energy tensor in
(\ref{efa2aa}) can be also written in the form \bea\label{efa2aaa}
\Pi^{\lambda\mu\nu}=-\frac{ie^2}{2}\int
\frac{d^4p}{(2\pi)^4}\;T^{\mu\lambda\nu}(p), \eea with \bea
&&T^{\mu\lambda\nu}(p)={\rm
tr}\bigl[W_{1}P_{R}\gamma^{\mu}W_{1}P_{R}\gamma^{\lambda}W_{1}P_{R}\gamma^{\nu}+W_{1}P_{R}\times\nonumber\\&&\gamma^{\mu}W_{1}P_{R}\gamma^{\lambda}W_{2}P_{L}\gamma^{\nu}+W_{1}P_{R}\gamma^{\mu}W_{2}P_{L}\gamma^{\lambda}W_{1}P_{R}\gamma^{\nu}+\nonumber\\&&W_{1}P_{R}\gamma^{\mu}W_{2}P_{L}\gamma^{\lambda}W_{2}P_{L}\gamma^{\nu}+W_{2}P_{L}\gamma^{\mu}W_{2}P_{L}\gamma^{\lambda}W_{2}\times\nonumber\\&&P_{L}\gamma^{\nu}+W_{2}P_{L}\gamma^{\mu}W_{2}P_{L}\gamma^{\lambda}W_{1}P_{R}\gamma^{\nu}+W_{2}P_{L}\gamma^{\mu}W_{1}\times\nonumber\\&&P_{R}\gamma^{\lambda}W_{2}P_{L}\gamma^{\nu}+W_{2}P_{L}\gamma^{\mu}W_{1}P_{R}\gamma^{\lambda}W_{1}P_{R}\gamma^{\nu}\bigl].
\eea Thus, taking into account the fact that
$\{\gamma_{5},\gamma^{\mu}\}=0$ and $(\gamma_5)^2=1$ and applying
the following relation for trace ${\rm
tr}(\gamma^{\lambda}\gamma^{\mu}\gamma^{\nu}\gamma^{\rho}\gamma_5)=4i\epsilon^{\lambda\mu\nu\rho}$,
we can write down the simple expression for self-energy tensor
$\Pi^{\mu\nu\lambda}$: \bea\label{proj}
&&\Pi^{\mu\nu\lambda}=ie^2\,\epsilon^{\lambda\mu\nu\rho}\int\frac{d^4p}{(2\pi)^4}
\frac{1}{(p^2-b^2)^4}\times\nonumber\\&&b_{\rho}\bigl((p+b)^{4}\!+\!(p-b)^{4}\bigl)-p_{\rho}\bigl((p+b)^{4}\!-\!(p-b)^{4}\!\bigl).
\eea
Now, we can write our effective action in the form
\bea\label{efa2aaaa} S_{eff}^{(2)}[b,A]=\frac{1}{2}\,\int d^{\,4}x
\,\epsilon^{\lambda\mu\nu\rho}k_{\rho}F_{\lambda\mu}A_{\nu}, \eea
where the $k_{\rho}$-parameter can be expressed in the form
\bea\label{tp1}
k_{\rho}&=&2ie^{2}\!\int\!\frac{d^{4}p}{(2\pi)^{4}}\Big[\frac{(p^{2}-b^{2})b_{\rho}-4(b\cdot
p)p_{\rho}}{(p^{2}-b^{2})^{3}}+\nonumber\\&&\frac{4\bigl[(p^{2}b^{2}+(b\cdot
p)^{2})b_{\rho}-2b^{2}(b\cdot
p)p_{\rho}\bigl]}{(p^{2}-b^{2})^{4}}\Big]. \eea Note that by power
counting, the momentum integral in (\ref{tp1}) involves finite terms
and terms with logarithmic divergence.

\section{Regularization schemes}
\label{sec4}

In the following we shall follow three different regularization
schemes: $i)$ the dimensional regularization, $ii)$ the momentum
cut-off regularization and a new one, $iii)$ the {\it temperature
regularization}. As our results show, by using any of these schemes
we got the same result for the coefficient of the induced
Chern--Simons-like term. This implies that in an extended massless
QED the induced Chern--Simons-like term is finite and determined
under these regularization schemes.

\subsection{Dimensional regularization}

In this subsection, we shall first adopt a simpler regularization
scheme such as dimensional regularization \cite{thooft} in order to
calculate the divergent integral. By assuming the usual statements
for this case, we change the number of dimensions from $4$ to $D$,
and we also change $\int d^{4}p/(2\pi)^{4}$ to
$(\mu^{2})^{\epsilon/2}\big[\int d^{D}p/(2\pi)^{D}\big]$, where
$\mu$ is an arbitrary parameter that identifies the mass scale with
$\epsilon=4-D$. The potential divergences in the momentum
integration come from the first term of the expression (\ref{tp1}).
In order to carry out the calculations, we use the following
relations \bea \int
\frac{d^{D}p}{(2\pi)^{D}}\frac{1}{(p^{2}-m^{2})^{2}}=\frac{i\,
\Gamma(\epsilon/2)}{(4\pi)^{D/2}(b^{2})^{\epsilon/2}}, \eea and \bea
\int
\frac{d^{D}p}{(2\pi)^{D}}\frac{p_{\mu}p_{\nu}}{(p^{2}-m^{2})^{3}}=\frac{i
\delta_{\mu\nu}}{4(4\pi)^{D/2}}\frac{\Gamma(\epsilon/2)}{(b^{2})^{\epsilon/2}},
\eea in (\ref{tp1}) so that the logarithmic divergence disappears.
The resulting induced Chern--Simons-like coefficient is finite and
exactly given by the value \bea\label{c1}
k_{\rho}=\frac{e^{2}}{8\pi^{2}}\,b_{\rho}. \eea


\subsection{Momentum {cut-off}-$\Lambda$ regularization}

To develop calculations via momentum {cut-off}, we consider the
expression (\ref{tp1}) and change the (3+1)-dimensional Minkowski
spacetime to a four-dimensional Euclidean space by performing the
Wick rotation $x_{0}\to-ix_{0}$, $p_{0}\to ip_{0}$, $b_{0}\to
ib_{0}$, $d^{4}x\to -id^{4}x$ and $d^{4}p\to id^{4}p$. Thus, we have
\bea\label{tp2}
&&k_{\rho}=-\frac{2ie^{2}}{(2\pi)^{4}}\int^{+\infty}_{-\infty}\times\nonumber\\&&\,d^{3}p\int^{+\infty}_{-\infty}d
p_{0}\Big[\frac{(p^{2}-b^{2})b_{\rho}-4(b\cdot
p)p_{\rho}}{(p^{2}-b^{2})^{3}}+\nonumber\\&&\frac{4\bigl[(p^{2}b^{2}+(b\cdot
p)^{2})b_{\rho}-2b^{2}(b\cdot
p)p_{\rho}\bigl]}{(p^{2}-b^{2})^{4}}\Big]. \eea 
For the sake of simplicity, in (\ref{tp2}) we shall choose only the
time-like component of our Lorentz-symmetry violating parameter,
such that $b_{\rho}=(b_{0},0)$, being  $b_{0}$ a nonzero component.
Such a simplification does not affect the four-vector coefficient of
the induced Chern--Simons-like term. It simplifies our calculations
without any loss of generality.

Taking into account the considerations above, we now write the
expression (\ref{tp2}) in spherical coordinates as \bea\label{tp3}
&&k_{0}=-\frac{ie^{2}b_{0}}{2\pi^{3}}\int_{0}^{\infty}\,d \vec{p}
\,\vec{p}^{2}\int^{+\infty}_{-\infty}d
p_{0}\times\nonumber\\&&\Big[\frac{\vec{p}^{2}-3p^{2}_{0}-b^{2}_{0}}{(\vec{p}^{2}
+p_{0}^{2}-b_{0}^{2})^{3}}+\frac{4b_{0}^{2}\vec{p}^{2}}{(\vec{p}^{2}+p_{0}^{2}-b_{0}^{2})^{4}}\Big].
\eea The integrals over $p_{0}$ are finite and can be directly
calculated. Hence, we have \bea
k^E_{0}&=&-\frac{5i}{8}\frac{e^{2}\,b_{0}}{\pi^{2}}\int_{0}^{\infty}\,d\vec{p}\frac{b_{0}^{2}p^{4}}{(\vec{p}^{2}
-b_{0}^{2})^{7/2}},\nonumber\\&=&-\frac{5i}{8}\frac{e^{2}\,b_{0}}{\pi^{2}}\int_{0}^{u=\Lambda/b_{0}}
du\,\frac{u^{4}}{(u^{2}-1)^{7/2}},\nonumber\\&=&\frac{ie^{2}\,b_{0}}{8\pi^{2}}
\frac{1}{(1-\frac{b_{0}^{2}}{\Lambda^{2}})^{7/2}}. \eea Therefore,
taking the limit $\Lambda \rightarrow \infty$, we find (in Euclidean
coordinates) \bea\label{c3}
k^{E}_{0}\to\frac{ie^{2}}{8\pi^{2}}\,b_{0}. \eea This result
coincides with the coefficient (\ref{c1}), previously obtained via
dimensional regularization.
\section{Chern--Simons coefficient induced under finite temperature effects}
\label{temp}
The effect of finite temperature in the context of violating Lorentz symmetry
has generated interesting studies in the literature
\cite{bpp,N3,Cervi:2001fg,Ebert:2004pq,Gomes:2007rv}.
In this case, let us assume
that the system is in the state of the thermal equilibrium with a
temperature $T=1/\beta$. Therefore, we make the following
substitutions: $p_{0}\equiv \omega_{0}=(n+1/2)\frac{2\pi}{\beta}$
and $(1/2\pi)\int d p_{0}\rightarrow \frac{1}{\beta}\sum_{n}$, where
$\omega_{0}$ is the Matsubara frequency for fermions. Again, we
consider the expression (\ref{tp1}) and rewrite it in the framework
of the imaginary time formalism as \bea\label{T1}
k_{\rho}(\beta)&=&\frac{-2i\,e^2}{\beta}\!\!\sum^{\infty}_{n=-\infty}
\!\!\int\frac{d^{3}\vec{p}}{(2\pi)^{3}}\Big[\frac{(p^{2}-
b^{2})b_{\rho}-4(b\cdot p)p_{\rho}}{(p^{2}-
b^{2})^{3}}+\nonumber\\&&\frac{4\bigl[(p^{2}b^{2}+(b\cdot
p)^{2})b_{\rho}-2b^{2}(b\cdot p)p_{\rho}\bigl]}{(p^{2}-
b^{2})^{4}}\Big]. \eea We can write the expression (\ref{T1}) for
the time-like and space-like components of the Lorentz-symmetry
violation parameter \bea\label{T2}
k_{0}(\beta)&=&\frac{2i\,e^2\,b_{0}}{\beta}\,\int\frac{d^{3}\vec{p}}{(2\pi)^{3}}
\,\sum^{\infty}_{n=-\infty}\Bigl[\frac{3}{(\omega_{0}^{2}+\vec{p}^{2}-b^{2}_{0})^{2}}
-\nonumber\\&&\frac{4\big(\vec{p}^{2}-b^{2}_{0}\big)}{(\omega_{0}^{2}+\vec{p}^{2}-b^{2}_{0})^{3}}
-\frac{4b^{2}_{0}\vec{p}^{2}}{(\omega_{0}^{2}+\vec{p}^{2}-b^{2}_{0})^{4}}\Bigl],
\eea and \bea\label{T21}
k_{i}(\beta)&=&\frac{-2i\,e^2}{\beta}\,\int\frac{d^{3}\vec{p}}{(2\pi)^{3}}
\times\nonumber\\&&\sum^{\infty}_{n=-\infty}\Bigl[\frac{b_{i}}{(\omega_{0}^{2}+\vec{p}^{2}
-b^{2}_{i})^{2}}+\frac{4\big(b_{i}^{3}\!-\!(\vec{b}\cdot\vec{p})^{2}p_{i}\big)}{(\omega_{0}^{2}
\!+\!\vec{p}^{2}\!-\!b^{2}_{i})^{3}}
\!+\nonumber\\&&\frac{4\big[\big(b_{i}^{4}+(\vec{b}\cdot\vec{p})^{2}\big)b_{i}
\!-\!2b_{i}^{2}(\vec{b}\cdot\vec{p})p_{i}\big]}{(\omega_{0}^{2}\!+\!\vec{p}^{2}\!-\!b^{2}_{i})^{4}}\Bigl].
\eea At this point, we can observe that both the 3-momentum
integrals and the sum over the Matsubara frequencies in (\ref{T2})
and (\ref{T21}) are convergent. This fact is interesting to our
investigations from two points of view. The first, of course,
concerns the study of high temperature effects on the parameter
$k_\mu$. The second, and new one, is the possibility of
understanding the temperature itself as a regulator of divergences.
It means the inclusion of finite temperature in the theory can be
also considered as a regularization scheme in the limit
$\beta\to\infty\ (T\to0)$. This is what we shall discuss just below.

\subsection{Temperature regularization}

$\bullet$ {The time-like component}

\begin{figure}[ht]\centerline{\includegraphics[{angle=90,height=6.0cm,angle=270,width=8.0cm}]
{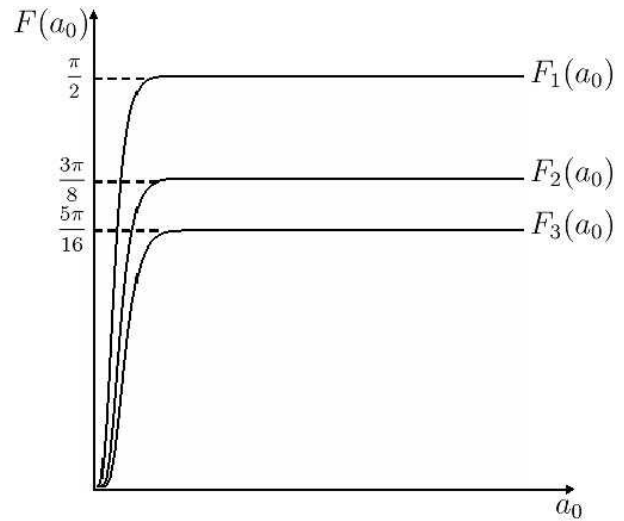}} \caption{The functions $F_1({a_0}), F_2({a_0})$, and
$F_3({a_0})$. At zero temperature ($\beta\to\infty$,
$a_0\to\infty$), they tend to distinct values, $\frac{\pi}{2}$,
$\frac{3\pi}{8}$, and $\frac{5\pi}{16}$,
respectively.}\label{passos1}
\end{figure}

To calculate the time-like component, we consider the expression
(\ref{T2}) and rewrite it in the following form \bea\label{T22}
&&k_{0}(\beta)\!=\!\frac{i\,e^2\,b_{0}}{\pi}\int\frac{d^{3}\vec{p}}{(2\pi)^{3}}
\sum^{\infty}_{n=-\infty}\Bigl[\frac{3a_{0}^{3}}{B_{0}^{3}
\big((n+\frac{1}{2})^{2}+a_{0}^{2}\big)^{2}}-\nonumber\\&&\frac{4a_{0}^{5}}{B_{0}^{3}
\big((n+\frac{1}{2})^{2}+a_{0}^{2}\big)^{3}}
-\frac{4a_{0}^{7}b^{2}_{0}\vec{p}^{2}}{B_{0}^{7}\big((n+\frac{1}{2})^{2}+a_{0}^{2}\big)^{4}}\Bigl],
\eea where $a_{0}=B_{0}\beta/2\pi$ and
$B_{0}=\big(\vec{p}^{2}-b^{2}_{0}\big)^{\frac{1}{2}}$. The series in
the above expression can be easily summed over the Matsubara
frequencies \cite{series} to give \bea\label{T221}
k_{0}(\beta)&=&\frac{i\,e^2\,b_{0}}{\pi}\,\int\frac{d^{3}\vec{p}}{(2\pi)^{3}}\times\nonumber\\&&\Bigl[\frac{3
F_{1}(a_{0})-4
F_{2}(a_{0})}{\big(\vec{p}^{2}-b_{0}^{2}\big)^{\frac{3
}{2}}}-\frac{4b^{2}_{0}\vec{p}^{2}F_{3}(a_{0})}{\big(\vec{p}^{2}-b_{0}^{2}\big)^{\frac{7}{2}}}\Bigl],
\eea where the functions $F_{1}(a_{0})$, $F_{2}(a_{0})$ and $F_{3
}(a_{0})$ are given by \bea\label{s1} F_{1}(a_{0})=\frac{\pi}{2} \,
\Big[ \tanh ( a_{0}\pi ) -a_{0}\pi \, \text{sech}^{2}( a_{0}\pi )
\Big]\,, \eea \bea\label{s2} F_{2}(a_{0})&=&\frac{\pi}{8} \, \Big[3
\tanh ( a_{0}\pi ) -a_{0}\pi \, \text{sech}^{2}( a_{0}\pi )\times\nonumber\\&&\big(3 +
2a_{0}\pi \tanh ( a_{0}\pi) \big) \Big], \eea and \bea\label{s3}
F_{3}(a_{0})&=&\frac{\pi}{48} \, \Big[15 \tanh (a_{0}\pi)+ a_{0}\pi
\text{sech}^{2}(a_{0}\pi)\times\nonumber\\&&\big(6a_{0}^{2}\pi^{2}\text{sech}^{2}(a_{0}\pi)\!-\!12a_{0}\pi\tanh
(a_{0}\pi)\!-\nonumber\\&&4 a_{0}^{2}\pi^{2}\!-\!15\big)\Big]. \eea Let us now consider
the limit $T\to0$. In this limit $a_{0}\propto\beta\to\infty$ and
the functions above achieve $F_{1}(a_{0})=\frac{\pi}{2}$,
$F_{2}(a_{0})=\frac{3\pi}{8}$ and $F_{3}(a_{0})=\frac{5\pi}{16}$, as
depicted in Fig.~\ref{passos1}. Now the expression (\ref{T221}) can
be written as \bea\label{T222}
k_{0}=\frac{-5i\,e^2\,b_{0}}{4}\,\int\frac{d^{3}\vec{p}}{(2\pi)^{3}}\frac{b^{2}_{0}\vec{p}^{2}}{\big(\vec{p}^{2}-b_{0}^{2}\big)^{\frac{7}{2}}}.
\eea This integral is manifestly finite and does not require any
further scheme of regularization. The exact value obtained from the
expression (\ref{T222}) is precisely the same result as that found
in (\ref{c1}) and (\ref{c3}) via dimensional and momentum {cut-off}
regularization in the absence of temperature.

\noindent $\bullet$ {The space-like component}

To calculate the space-like component, we proceed just as in the
earlier case. We consider the expression (\ref{T21}) and rewrite it
in the following form \bea\label{T2222}
k_{i}(\beta)&=&\frac{-i\,e^2}{\pi}\,\int\frac{d^{3}\vec{p}}{(2\pi)^{3}}\,\sum^{\infty}_{n=-\infty}\times\nonumber\\&&\Bigl[\frac{b_{i}a_{i}^{3}}{B_{i}^{3}\big((n+\frac{1}{2})^{2}+a_{i}^{2}\big)^{2}}+\frac{\big[b_{i}^{3}-(\vec{b}\cdot\vec{p})p_{i}\big]a_{i}^{5}}{B_{i}^{5}\big((n+\frac{1}{2})^{2}+a_{i}^{2}\big)^{3}}+\nonumber\\&&\frac{4\big[\big(b_{i}^{4}+(\vec{b}\cdot
\vec{p})^{2}\big)b_{i}-2b_{i}^{2}(\vec{b}\cdot\vec{p})p_{i}\big]a_{i}^{7}}{B_{i}^{7}
\big((n+\frac{1}{2})^{2}+a_{i}^{2}\big)^{4}}\Bigl], \eea where
$a_{i}=B_{i}\beta/2\pi$ and
$B_{i}=\big(\vec{p}^{2}-b^{2}_{i}\big)^{\frac{1}{2}}$. Here, just as
in the time-like case, the sum over the Matsubara frequencies can be
directly calculated to give \bea\label{T2222.1}
k_{i}(\beta)&=&\frac{-i\,e^2}{\pi}\,\int\frac{d^{3}\vec{p}}{(2\pi)^{3}}\times\nonumber\\&&\Bigl[\frac{b_{i}
F_{1}(a_{i})}{\big(\vec{p}^{2}-b^{2}_{i}\big)^{\frac{3}{2}}}+\frac{4\big[b_{i}^{3}-(\vec{b}\cdot\vec{p})p_{i}\big]F_{2}(a_{i})}{\big(\vec{p}^{2}-b^{2}_{i}\big)^{\frac{5}{2}}}+\nonumber\\&&\frac{4\big[\big(b_{i}^{4}\!+\!(\vec{b}\cdot
\vec{p})^{2}\big)b_{i}-2b_{i}^{2}(\vec{b}\cdot\vec{p})p_{i}\big]
F_{3}(a_{i})}{\big(\vec{p}^{2}-b^{2}_{i}\big)^{\frac{7}{2}}}\Bigl].
\eea The functions $F_{1}(a_{i})$, $F_{2}(a_{i})$ and $F_{3}(a_{i})$
have the same functional form given by the expressions (\ref{s1}),
(\ref{s2}) and (\ref{s3}). Now we can use the identity
$p_{i}p_{j}=\frac{\vec{p}^{2}}{3}\,\delta_{ij}$ in (\ref{T2222.1}),
to obtain the simpler equation \bea\label{T2222.12}
k_{i}(\beta)&=&\frac{-ie^2
b_{i}}{2}\,\int\frac{d^{3}\vec{p}}{(2\pi)^{3}}\,\Bigl[
\frac{F_{1}(a_{i})-\frac{4}{3} F_{2}(a_{i})
}{\big(\vec{p}^{2}-b^{2}_{i}\big)^{\frac{3}{2}}}+\nonumber\\&&\frac{8}{3}\Big[\frac{b_{i}^{2}F_{2}(a_{i})}{\big(\vec{p}^{2}-b^{2}_{i}\big)^{\frac{5}{2}}}+\frac{b_{i}^{4}
F_{3}(a_{i})}{\big(\vec{p}^{2}-b^{2}_{i}\big)^{\frac{7}{2}}}\Big]\Bigl].
\eea By using again the limit $T\to0$, we have \bea\label{T2222.123}
k_{i}=-ie^2
b_{i}\int\frac{d^{3}\vec{p}}{(2\pi)^{3}}\,\Big[\frac{b_{i}^{2}}{\big(\vec{p}^{2}-b^{2}_{i}\big)^{\frac{5}{2}}}+\frac{5}{6}\frac{b_{i}^{4}
}{\big(\vec{p}^{2}-b^{2}_{i}\big)^{\frac{7}{2}}}\Bigl]. \eea In this
case we also obtain only finite integrals that does not require any
further scheme of regularization. Thus, similarly to what happen
with the time-like component, the exact value obtained from the
expression (\ref{T2222.123}) is the same as the result found in
(\ref{c1}) and (\ref{c3}) in the absence of temperature.

In summary, we find that the induced Chern--Simons-like coefficient
that is {\it temperature regularized} has the following time-like
and space-like components \bea
k_{0}=\frac{ie^2}{8\pi^2}\,b_0,\;\;
k_{i}=\frac{ie^2}{8\pi^2}\,\;b_{i},\;\;\!\!\!\! (a_0\to\infty
\mbox{ or } T\to 0).
\eea



\section{Conclusions}
\label{conc}

In this paper we have shown that the induced Chern--Simons-like term
in an extended {\it massless} QED is finite and determined at least
in the framework of the three regularization schemes we have used:
$i)$ the dimensional regularization, $ii)$ the momentum cut-off
regularization and a new one, $iii)$ the {\it temperature
regularization}. A similar observation concerning determined
Chern--Simons-like term in extended massless QED has also been done
in the Ref.~\cite{8}. We state here that this has to do with the
fact that one can relate an extended masless QED in 3+1 dimensions
with a massive QED in 2+1 dimensions where the induced Chern--Simons
term is finite and determined since the theory is finite. By
dimensional reduction from 3+1 down to 2+1 dimensions, one component
of the 3+1 Chern--Simons-like term gives the coefficient of the 2+1
Chern--Simons term. Since this term comes from a finite theory, it
should be a finite and determined term. This automatically implies
that at least one component of the four-vector coefficient $k_\mu$
of the 3+1 Chern--Simons-like term, i.e. the 2+1 Chern--Simons
counterpart, is also finite and determined. Although this fact may
guarantee that only one component is well determined, in our
explicit calculations, in order to induce the Chern--Simons-like
term, we have found that all components are finite and well
determined.

{\acknowledgments} We would like to thank CNPq,
CAPES, PNPD/PRO\-CAD-CAPES, and FAPERJ for partial financial support.


\end{document}